\documentclass[11pt]{article}
\usepackage{amssymb, bm,amsmath,amsthm,amsfonts,a4,bezier}
\usepackage{hyperref}
\usepackage{amscd}
\usepackage{graphpap}
\usepackage[pdftex]{color,graphicx,xcolor}
\usepackage{enumerate}
\usepackage{setspace}
\usepackage{comment}
\usepackage{mathtools}
\usepackage{indentfirst}
\usepackage{verbatim,bbm}
\usepackage{extarrows}
\usepackage[pdftex]{color}
\usepackage{relsize}
\usepackage{calrsfs}
\usepackage{tikz}
\usepackage{float}
\usepackage[all]{xy}
\usepackage{mathrsfs}
\usetikzlibrary{patterns}

\DeclareGraphicsExtensions{.jpg,.pdf,.eps,.png}
\newtheorem{theorem}{Theorem}

\renewcommand{\leq}{\leqslant}
\renewcommand{\geq}{\geqslant}

\newcommand{\Z}{{\mathbb{Z}}}

\renewcommand{\L}{{\Lambda}}

 \let\b=\beta   
 \let\g=\gamma \let\h=\eta     
\let\m=\mu  \let\o=\omega     
 \let\s=\sigma  
  
  \let\G=\Gamma \let\L=\Lambda 
\let\O=\Omega

\renewcommand{\o}{{\omega}}

\title{Anisotropic Ising model in $d+s$ dimensions}

\begin{document}
\author{
    %newline for better readability of affilations
    Estevão F. Borel\thanks{Departamento de Matemática, Universidade Federal de Minas Gerais, Belo Horizonte, Brazil. \newline},
    Aldo Procacci$^*$,
    Rémy Sanchis$^*$,
    Roger W. C. Silva\thanks{Departamento de Estatística, Universidade Federal de Minas Gerais, Belo Horizonte, Brazil.\newline}
}
\maketitle

%\baselineskip=15pt
%\vskip 10mm
%\centerline{\today}
%\vskip 10mm

\begin{abstract} In this note, we consider the asymmetric nearest neighbor ferromagnetic Ising model on the $(d+s)$-dimensional unit cubic lattice
$\Z^{d+s}$, at inverse temperature $\beta=1$ and with coupling constants $J_s>0$ and $J_d>0$ for edges of $\Z^s$ and $\Z^d$, respectively. We obtain
a lower bound for the critical curve in the phase diagram of $(J_s,J_d)$. In particular, as $J_d$ approaches its critical value from below, our
result is directly related to the so-called dimensional crossover phenomenon.  \\


\end{abstract}

\onehalfspacing

\section{Introduction}

In this paper, we consider an asymmetric nearest neighbor ferromagnetic Ising model in the $(d+s)$-dimensional unit cubic lattice
$\mathbb{Z}^{d+s}$, with coupling constants $J_d>0$ and $J_s>0$ in the hyperplanes  $\mathbb{Z}^{d}$ and  $\mathbb{Z}^{s}$, respectively.
\begin{comment}
in which the coupling constant is $J_d$ in the $\mathbb{Z}^{d}$ hyperplane and
$J_s<J_d$ in the  $\mathbb{Z}^{s}$ hyperplane.
\end{comment}
Anisotropic lattice spin systems have been the subject of great interest within the physics community since the sixties.
The study of such models  has been tackled both numerically, mainly via Monte Carlo simulations (see
e.g. \cite{KKB,KASL, L, Yu} and references therein), and theoretically, via mean-field, Bethe approximation, truncated high-temperature
expansion of the susceptibility, etc...  (see e.g. \cite{LS1, LS2, NJ,OE, Su, Ya, Yu, ZSH}).  A  strong motivation to study anisotropic
systems is to investigate finite-size effects in realistic materials modeled by quasi-two-dimensional (thin films) and quasi-one-dimensional
spin systems. Furthermore, exploring these systems could provide valuable insights into isotropic models, notably the three-dimensional
Ising model (see. e.g.  \cite{VPRL} and references therein).

Rigorous results on the asymmetric Ising model on $\mathbb{Z}^{d+s}$ have been obtained mainly in the case $\mathbb{Z}^{1+s}$ with strong coupling
in one dimension and small coupling in the remaining directions. In particular, in a well-known
 article (see \cite{F}),  Fisher derived an asymptotic bound on the critical temperature of the $\mathbb{Z}^{1+s}$ anisotropic Ising model in the
 limit $\frac{J_s}{J_1}\to 0$. It has also  been shown rigorously (see \cite{MPS}) that the free energy of the Ising model on
  $\mathbb{Z}^{1+s}$ is analytic for any inverse temperature $\b$ if $J_s$ is small enough (depending on $J_1$ and the inverse temperature).   These
  rigorous results rely heavily on the fact that when $d=1$ the one-dimensional system is in the gas phase at all temperatures so that the standard
  high-temperature expansion can be used with effectiveness. On the other hand, to obtain the same kind of results of references \cite{F,MPS}  in
  the case $\mathbb{Z}^{d+s}$, $d\ge 2$, is expected to be trickier since the $d-$dimensional system exhibits a phase transition and the usual
  high-temperature expansion turns to be much more difficult to control.

In this paper, we somehow extend the results obtained in \cite{F} and \cite{MPS} for the case $\mathbb{Z}^{1+s}$ to the case $\mathbb{Z}^{d+s}$ with
$d\ge 1$. Namely, we show that for all $J_d$  below the critical reduced temperature $J_d^c$, the susceptibility of the $(d+s)$-dimensional system
is finite when the coupling $J_s$  is sufficiently small (inversely proportional to the susceptibility $\chi(J_d)$ of the $d$-dimensional system).
A similar result was obtained by two of us in  \cite{SS}   for the Bernoulli anisotropic bond percolation model on $\mathbb{Z}^{d+s}$,  in which
edges in the $\mathbb{Z}^{d}$ hyperplane are open with probability $p<p_c(d)$ and edges parallel in the $\mathbb{Z}^{s}$ hyperplane are open with
probability $q$. In \cite{SS},   probabilistic arguments were applied and in particular, a crucial use of the van den Berg–Kesten (BK) inequality
has been made.
In the present paper, to get the analogous result for the anisotropic Ising model in $\mathbb{Z}^{d+s}$ (for which
BK inequalities are not available), we use an alternative (w.r.t. the high
temperature) expansion  of the Ising partition function, namely the so-called random current representation. This powerful technique,  introduced
in the eighties  by Aizenman \cite{A}, was  widely used by several authors in the following decades, and recently it has employed  as a crucial tool
in several remarkable papers, e.g., \cite{AD}, \cite{ADS}, \cite{ADTW}.

As mentioned earlier, our results may be of interest in the study of realistic quasi-two-dimensional   magnets which can be modeled by a two-dimensional
sub-critical Ising bilayer.  When the interaction between the two two-dimensional layers is weak and the transverse interaction is sub-critical, the
bilayer system is expected to still exhibit sub-critical behavior. However, as the coupling between the layers strengthens, the overall system may
exhibit spontaneous magnetization. Our result implies rigorously that as long as the inter-layer interaction between layers is below the inverse
two-dimensional susceptibility (with a constant factor 1/(2s)), the global system remains sub-critical.
The understanding of how several layers of  2D slightly sub-critical systems with small interactions between them can start to behave as a
$(2+s)$-dimensional system, is the so-called dimensional crossover phenomenon (see e.g. Sec. VI in \cite{LS2}). This phenomenon is characterized by
a critical exponent, which is believed to depend on the original dimension of the layers, but not on the target dimension. Moreover, numerical
simulations and formal calculations (see e.g. \cite{LS2,NJ}), suggest that this critical exponent equals the exponent of the susceptibility of the
original dimension. Our results imply rigorously an inequality between the two exponents.

\section{The model and results}\label{bkgr}

Let $\Z^{d+s}=\Z^d\times \Z^s$ be the $(d+s)$-dimensional unit cubic lattice.
We will denote by $\mathbb{E}^{d+s}$ the set of nearest neighbor pairs of $\mathbb{Z}^{d+s}$ so that $\mathbb{G}^{d+s}$ is the graph with vertex set
$\mathbb{Z}^{d+s}$ and edge set $\mathbb{E}^{d+s}$. Given two vertices $x,y\in \mathbb{Z}^{d+s}$, we denote by $|x-y|$ the usual graph distance
between $x$ and $y$  (i.e. the edge-length of the shortest path between $x$ and $y$). We will suppose that $\mathbb{Z}^{d+s}$  is equipped with the
usual operation of sum.
We represent hereafter  a site $x\in\mathbb{Z}^{d+s}$ as $x=(u,t)$, where $u\in\mathbb{Z}^{d}$ and $t\in\mathbb{Z}^{s}$.

Given an integer  $N$, we denote by  $\Lambda_N\subset \mathbb{Z}^{d+s}$ the hypercube with side length $2N+1$, centered at the origin, so that
$\L_N\to \infty$ means that $N\to \infty$.  We denote by $E_N$ the set of edges of $\mathbb{E}^{d+s}$ with both endpoints in $\L_N$,
so that  $\mathbb{G}^{d+s}|_{\L_N}=(\L_N, E_N$) is the restriction of $\mathbb{G}^{d+s}$ to $\L_N$. Note that
$\L_N=\bar\L_N\times\hat\L_N$ where $\bar\L_N$  denotes  the $d$-dimensional hypercube in $\Z^d$ of size $2N+1$ centered at the origin and  $\hat\L_N$ denotes the $s$-dimensional hypercube in $\Z^s$  of size $2N+1$ centered at the origin.
 Given $w\in \hat\L_N$, we set
 $\Lambda^w_{N}=\{(u,t)\in \Z^{d+s}: t=w\}$. Namely $\Lambda^w_{N}$ is the subset of $\Lambda_N$ formed by sites of $\L_N$ with $w$ as the second
 coordinate. Similarly, $E^w_N$ will denote the set of edges with both endpoints in $\L_N^w$.
Observe that $\L_N^w$ is a $d$-dimensional box of side length  $2N+1$  centered at $(0,w)$.

To each vertex  $x\in \L_N$ we associate a random variable $\s_x$ taking values in the set $\{+1, -1\}$.
A spin configuration in $\L_N$ is a function $\bm\s: \L_N\to \{+1, -1\}:x\mapsto \s_x$.
The energy  of a configuration $\bm \s$ is given by the (free boundary condition) Hamiltonian
\begin{equation*}
H_{\L_N}(\bm \s)= -\sum_{\{x,y\}\in E_N}J_{\{x,y\}}\s_x\s_y,
\end{equation*}
where
$$
J_{\{x,y\}}=\begin{cases}
J_s & {\rm if}\;\;x=(u,t)\;\;{\rm and}\;\; y=(u, t')\;\; {\rm with}\;\;|t-t'|=1,\\
J_d & {\rm if}\;\;x=(u,t)\;\;{\rm and}\;\; y=(u', t)\;\; {\rm with}\;\;|u-u'|=1,
\end{cases}
$$
with $J_s>0$ and $J_d>0$.  In what follows, an edge $\{x,y\}\in E_N$ is called \textit{vertical} if $x=(u,t)$, $y=(u, t')$  with $|t-t'|=1$, and
called \textit{planar} if $x=(u,t)$, $y=(u', t)$  with $|u-u'|=1$.  So $J_{\{x,y\}}=J_s$ if $\{x,y\}$ is vertical
 and $J_{\{x,y\}}=J_d$ if $\{x,y\}$ is planar.
\begin{comment}
The partition  function and the two-point function of the system at inverse temperature $\b>0$  are given respectively by
\end{comment}

The partition function  of the system   is given by
\begin{equation*}
Z_{\L_N}(J_d,J_s)= \int d\m_{\L_N}(\bm \s)
e^{-H_{\L_N}(\bm \s)}=\int d\m_{\L_N}(\bm \s)\prod_{b\in E_N}e^{J_b\s_x\s_y},
\end{equation*}
where  $\int d\m_{\L_N}(\bm \s)$ is a short notation for $\prod_{x\in \L_N}\frac{1}{2}\sum_{\s_x=\pm1}$ (a product probability measure). Moreover,
without loss of generality, we have set the inverse temperature $\b=1$.

The two-point correlation function of the $(d+s)$-system is then defined as
\begin{equation}\label{two}
\langle \sigma_x \sigma_y \rangle_{\L_N}=
\frac{\int  d\m_{\L_N}(\bm \s)\s_x\s_y
e^{- H_{\L_N}(\s_\L)}}{Z_{\L_N}(J_d,J_s)}
=\frac{\int  d\m_{\L_N}(\bm \s)\s_x\s_y\prod_{\{x,y\}\in E_N}e^{J_b\s_x\s_y}}{Z_{\L_N}(J_d,J_s)}.
\end{equation}
In general, for any  set $U\subset \L_N$, letting $E_U=\{\{x,y\}\in \L_N: \{x,y\}\subset U\}$, we set
$$
H_{U}(\bm s)= -\sum_{\{x,y\}\in E_U}J_{\{x,y\}}\s_x\s_y,
$$
$$
Z_{U}(J_d,J_s)= \int d\m_{\L_N}(\bm \s)
e^{-H_{U}(\bm \s)},
$$
and, for any $x,y\in U$,
$$
\langle \sigma_x \sigma_y \rangle_{U}=
\frac{\int  d\m_{\L_N}(\bm \s)\s_x\s_y
e^{- H_{U}(\s_\L)}}{Z_{U}(J_d,J_s)}.
$$
According to the above notations, for any $t\in \hat\L_N$, we have that
$$
\langle \sigma_x \sigma_y \rangle_{\L^t_N}= \frac{\int  d\m_{\L_N}(\bm \s)\s_x\s_ye^{- H_{\L^t_N}(\s_\L)}}{Z_{\L_N^t}(J_d,J_s)},
$$

The finite volume susceptibility  function of the system is defined as  \begin{equation*}
    \chi_{\L_N}(J_d,J_s)\coloneqq \sup_{x\in \L_N}\Bigg\{\sum_{y\in \L_N}\langle \sigma_x\sigma_y\rangle_{\Lambda_N}\Bigg\},
\end{equation*}
so that
\begin{equation}\label{kids}
 \chi_{d+s}(J_d,J_s)= \lim_{N\to \infty}\chi_{\L_N}(J_d,J_s)
\end{equation}
is the susceptibility of the anisotropic  $(d+s)$-dimensional Ising model.
%For more details on the Ising model, the interested reader can see \cite{FV}, for instance.

 For any $t\in \hat\L_N$, let
\begin{equation}\label{kidn}
\chi_{\L_N^t}(J_d)=\sup_{x\in \L_N^t}
\Bigg\{\sum_{y\in \Lambda_N^t}\langle \sigma_x\sigma_y\rangle_{\L_N^t}\Bigg\},
\end{equation}
 so that
\begin{equation}\label{kid}
\chi_d(J_d)=\lim_{N\to\infty}\chi_{\L_N^t}(J_d)
\end{equation}
 is the susceptibility of the $d$-dimensional Ising model with ferromagnetic interaction $J_d$.

 We are now in a position to state our main result.
\begin{theorem}\label{th1}
Take any $J_d$ such that

$$
\chi_d(J_d)<+\infty,
$$
and let $J_s$ such that
$$
\tanh(J_s)< \frac{1}{2s\chi_{d}(J_d)}.
$$
Then
$$
\chi_{d+s}(J_d, J_s)<+\infty.
$$
\end{theorem}

\noindent{\bf Remark}. As mentioned in the introduction,
Theorem \ref{th1} is related to the so-called dimensional crossover phenomenon.
%Loosely speaking, the term crossover is commonly used to describe how a transition between two separate phases occurs when parameters are
%appropriately tuned.
Denoting by  $J_d^c$  the critical inverse reduced temperature of the $d$-dimensional system, one can define,  for any  $J_d>0$, the function
$J_s^c(J_d):[0,\infty)\rightarrow [0,\infty)$, where
\begin{equation}\label{coup_crit}
J_s^c(J_d)=\sup\{J_s: \chi_{d+s}(J_d,J_s)<+\infty\}.
\end{equation}
By definition, $J_s^c(J_d)=0$ if $J_d\ge J^c_d$.
Theorem \ref{th1} determines a region in the $(J_d,J_s)$ plane where no phase transition occurs in
the $(d+s)$-system and it also gives an upper bound for the function $J_s^c(J_d)$ as $J_d$ varies in the interval $[0, J_d^c)$.
There is a strong interest in understanding the behavior  of the function  $J_s^c(J_d)$ when $J_d<J_d^c$, and in particular it is widely believed
that there exists a constant    $\phi_d>0$ such that
\[
J_s^c(J_d)\approx |J_d-J_d^c|^{\phi_d}, ~~{\rm as} ~~ J_d\uparrow J_d^c,
\]
where the symbol $\approx$ stands for log equivalence, that is, $f(x)\approx g(x)$ if $\displaystyle\lim_{\beta\uparrow \beta_c}\frac{\log
f(x)}{\log g(x)}=1$. The constant
 $\phi_d$ is the so-called crossover critical exponent.
On the other hand, the  $d$-dimensional Ising susceptibility $\chi_d(J_d)$ is known  to behave like
\begin{equation*}
\chi_d(J_d)\approx |J_d-J_d^c|^{-\g_d}, \mbox{ as }
 J_d\uparrow J_d^c,
\end{equation*}
where
$\g_d>0$ is the susceptibility $d$-dimensional critical exponent.
\begin{comment}
Actulally, for the Ising model, the critical exponent $\g_d$ has been determined exactly for $d=2$ (see \cite{CGN,O,Y}) and for $d\geq 4$ (see
\cite{A,AF,S}) while  for the case $d=3$ only numerical results are available.
\end{comment}
As an immediate corollary, Theorem \ref{th1} implies that $\phi_d\le \g_d$ for  all $d\ge 2$, and a lower bound for $J_s^c(J_d)$ still inversely
proportional to $\chi_d(J_d)$ would imply that $\phi_d=\g_d$.
The conjectured equality of the crossover critical exponent $\phi_d$ and susceptibility critical exponent $\g_d$ has been
discussed in the physics literature by several authors, see for instance \cite{Abe,Su, L, LS1,LS2}.

%On the 3D Ising model, one can consider a Hamiltonian with coupling constants $J$ %between nearest neighbor spins in the $xy$-plane and
%$J_z=\lambda J$ for nearest 
%neighbor spins in the $z$-direction. Many studies of the dimensional crossover have 
%suggested that the critical temperature behaves as $\lambda^{1/\phi}$,  with $\phi$ 
%been the crossover critical exponent. It is also suggested that $\phi$ is equal to 
%the susceptibility critical exponent when $\lambda=0$, but no mathematical proof has 
%been given so far. For more details see \cite{LS1,LS2} and references there in.

\section{The random current representation}\label{backbone}

As mentioned in the introduction, in order to prove Theorem \ref{th1} we will use the so-called \textit{random-current representation} for the Ising
model introduced by Aizenman in \cite{A}. We will describe this technique  here below following mainly reference \cite{AF}.

Let $\mathfrak{F}(E_N)$ be the set of all functions $\h: E_N\to \mathbb{N}:b\mapsto \h_b$. Then
we can expand the exponential inside the product and   rewrite the partition function $Z_{\L_N}(J_d,J_s)$ as
$$
Z_{\L_N}(J_d,J_s)= \sum_{\h\in \mathfrak{F}(E_N)}W(\h)\int  d\m_{\L_N}(\bm \s)\prod_{x\in \L_N} (\s_x)^{\sum_{b\ni x}\h_b},
$$
\begin{comment}
$$
Z_{\L_N}(J_d,J_s)= \sum_{\h\in \mathfrak{F}(E_N)}W(\h)\int  d\m_{\L_N}(\bm \s)\prod_{x\in \L_N} (\s_x)^{\sum_{b\ni x}\h_b}\h_b},
$$
\end{comment}
where we have denoted shortly 
$$
\sum_{b\ni x}\h_b=\sum_{b\in E_N\atop  x\in b}\h_b
$$ 
and we have set
$$
W(\h)=\prod_{b\in E_N} \frac{(J_b)^{\h_b}}{\h_b!}.
$$
Observe that the integral $I(\h)=\int d\m_{\L_N}(\bm \s)\prod_{x\in \L_N} (\s_x)^{\sum_{b\ni x}\h_b}$ is zero, unless
$\sum_{b\ni x}\h_b$ is even for all $x\in \L$, in which case $I(\h)=1$.
Hence
\begin{equation*}
Z_{\L_N}(J_d,J_s)= \sum_{\substack{\h\in \mathfrak{F}(E_N)\\\partial\h=\emptyset}}W(\h),
\end{equation*}
\begin{comment}
\begin{equation*}
Z_\L(J_d,J_s)= \sum_{\h\in \mathfrak{F}(E)\atop\partial\h=\emptyset}W(\h),
\end{equation*}
\end{comment}
where we have set
$$
\partial\h=\{x\in \L: \;\sum_{\substack{b\ni x}}\eta_b \;\mbox{is odd}\}.
$$
In general, given any $\h\in \mathfrak{F}(E_N)$,
the vertices in $\partial\h$ are called {\it sources} of $\h$, and if  $\partial\h=\emptyset$, then $\h$ is called {\it sourceless}.
Proceeding similarly we have that the  random-current expansion for the two-point function is

\begin{equation*}
   \langle \sigma_x \sigma_y \rangle_{\Lambda_N}=\sum_{\substack{\h\in \mathfrak{F}(E_N)\\\partial \eta=\{x,y\}}}\frac{W(\eta)}{Z_{\L_N}(J_d,J_s)}.
\end{equation*}

We now rewrite the function
$\langle \sigma_x \sigma_y \rangle_{\L_N}$ as a sum of edge-self-avoiding walks from $x$ to $y$.
Given an  edge $\{x,y\}\in E_N$, the ordered pair $(x,y)$ will be called a {\bf step} from $x$ to $y$.
For any $x\in \L_N$, we establish an arbitrary order (denoted by $\preceq$) for the set of steps emerging from $x$ (i.e. for those $(x,y)$ such
that
$|x-y|=1$). For each site $x$, and each {\it step} $(x,z)$, we consider the set $\Gamma_{(x,z)}$ formed by the edges $b=\{x,y\}$ such that
$(x,y)\preceq(x,z)$. This set will be referred to as the set of edges  {\it canceled} by $(x,z)$. In particular, since $(x,y)\preceq(x,y)$, a step
$\{x,y\}$ cancels itself.

We recall that a path in $\L_N$ is a sequence  $p=\{x_0,x_1,...,x_n\}$ of vertices of $\L_N$ such that $\{x_{i-1},x_i\}\in E_N$ for all $i=
1,\cdots,n$.
We say that a path $p=\{x_0,x_1,...,x_n\}$ is {\it consistent} if, for each $k=1,...,n$, we have that $\{x_{k-1},x_k\}\notin
\cup_{i=1}^{k-1}\Gamma_{(x_{i-1},x_i)}$. That is, each step used in this path is not associated with an edge that was canceled by the previous
steps. If $p=\{x_0,x_1,...,x_n\}$ is a consistent path, we denote by ${p}^*$ the set of all edges canceled by $p$, that is
$p^*=\cup_{i=1}^{n}\Gamma_{(x_{i-1},x_i)}$. Clearly, by construction, a consistent path is always edge-self-avoiding. We denote by
$C_{xy}(\L_N)$ the set of all consistent paths in $\L_N$  from $x$ to $y$.

We now define a function $\O$, which associates to each current configuration $\eta$ with
$\partial \h=\{x,y\}$, a consistent path $\o=\O(\h)$ from $x$ to $y$, which belongs to $C_{xy}(\L_N)$. As in \cite{AF}, such a consistent path will
be called the {\it backbone} of $\eta$.

Given  $\h$ with $\partial \h=\{x,y\}$, let $\G_\h$ be the set of edges $b\in E$ such $\h_b$ is odd.
Then, $\G_\h$ forms a subgraph of $(\L_N,E_N)$ (in general not connected) such that every vertex has degree either even or zero, except on $x$ and
$y$, whose degrees are odd.
The graph $\G_\h$ necessarily contain a connected component, say $\g_\h^{x,y}$, which contains $x$ and $y$. Therefore, we can look at this
connected component $\g_\h^{x,y}$ (seen as a set of edges in $E_N$), uniquely determined by $\h$, and associate to it a consistent path $\o=\O(\h)$. This is the
path $\o=\{z_0=x,z_1\},\{z_1,z_2\}, \dots, \{z_{k-1},z_k=y\}$ for some $k\ge |x-y|$ such that  for any $i=1,2,\dots, k$, $(z_{i-1},z_i)$ is the minimal step according to the order established
among the steps emerging from $z_{i-1}$  associated to edges of $\g_\h^{x,y}-\{z_0,z_1\}\cup\{z_1,z_2\}\cup\dots\cup\{z_{1-2},z_{i-1}\}$.

Once the function $\O$ is defined, we now can rewrite
$\langle \sigma_x \sigma_y \rangle_{\L_N} $ as
$$
\langle \sigma_x \sigma_y \rangle_{\L_N}= \sum_{\o\in C_{xy}(\L_N)}\sum_{\substack{\h\in \mathfrak{F}(E_N)\\\partial \eta=\{x,y\},
\O(\h)=\o}}\frac{W(\eta)}{Z_{\L_N}(J_d,J_s)}.
$$
Note that if $\O(\h)=\o$, then  $\h$ is odd on the edges of the set $\o$ and is even on the edges of $\o^*\setminus \o$. Also,
$\h$ restricted to $E_N\setminus \o$, as well as to $E_N\setminus \o^*$, is such that $\partial \h=\emptyset$. Therefore, setting
shortly $Z_N=Z_{\L_N}(J_d,J_s)$, we have that
$$
\begin{aligned}
\sum_{\substack{\h\in \mathfrak{F}(E_N)\\\partial \eta  =\{x,y\}, \O(\h)=\o}}\frac{W(\eta)}{Z_N}& =\prod_{b\in \o}\sinh( J_b)
\prod_{b\in \o^*\setminus \o}\cosh( J_b)\sum_{\substack{\h\in \mathfrak{F}(E_N\setminus \o^*)\\\partial \eta=\emptyset}}\frac{W(\eta)}{Z_N}\\
&= \prod_{b\in \o}\tanh(J_b)
\prod_{b\in \o^*}\cosh( J_b)\sum_{\substack{\h\in \mathfrak{F}(E_N\setminus \o^*)\\ \partial \eta=\emptyset}}\frac{W(\eta)}{Z_N}\\
&= \prod_{b\in \o}\tanh(J_b)\left[
\prod_{b\in \o^*}\cosh( J_b)\sum_{\substack{\h\in \mathfrak{F}(E_N\setminus \o^*)\\\partial \eta=\emptyset}}\frac{W(\eta)}{Z_N}\right]\\
&= \prod_{b\in \o}\tanh(J_b)
\sum_{\substack{\h\in \mathfrak{F}(E_N):\;\partial \eta=\emptyset \\ \h\;{\rm even\; on}\;\o^*}}
\frac{W(\eta)}{Z_N},
\end{aligned}
$$
where the last summation is over all sourceless current configurations $\eta$ on $E$ with the additional restriction that $\eta_b$ is even on all
edges $b$ canceled by $\omega$.
Hence, we can rewrite $\langle \sigma_x\sigma_y\rangle_\Lambda$  as
\begin{equation}\label{eq:backboneexp}
    \langle \sigma_x\sigma_y\rangle_{\L_N} =\sum_{\o\in C_{xy}(\L_N)}\rho_{E_N}(\omega),
\end{equation}
where
\begin{equation}\label{eq:rho}
    \rho_{E_N}(\omega)=\prod_{b\in \o}\tanh(J_b)
\sum\limits_{\substack{\h\in \mathfrak{F}(E_N):\;\partial \eta=\emptyset \\ \h\;{\rm even\; on}\;\o^*}}\frac{W(\eta)}{Z_N}.
\end{equation}
Observing that
$$
\sum\limits_{\substack{\h\in \mathfrak{F}(E_N):\;\partial \eta=\emptyset \\ \h\;{\rm even\; on}\;\o^*}}\frac{W(\eta)}{Z_N}
\le
\frac{\sum_{\h\in \mathfrak{F}(E_N):\,  \partial \eta=\emptyset}W(\h)}{Z_N}=1,
$$
we obtain straightforwardly the following upper bound
\begin{equation}\label{prot}
    \rho_E(\omega)\leq \prod_{b\in\omega}\tanh(J_b).
\end{equation}

\section{Proof of Theorem \ref{th1}}
To prove Theorem \ref{th1} we shall use two properties of the weights $\rho_{E_N}(\omega)$ defined in (\ref{eq:rho}). The interested reader can
check their proofs in Section 4.2 of \cite{AF}.

\begin{itemize}
\item[{\bf a)}]
Let $U\subset E_N$ be a set of edges of $\L_N$
and let and $\omega\subset U$ be a consistent path.  Then
\begin{equation}\label{proa}
\rho_{E_N}(\omega)\leq \rho_{U}(\omega).
\end{equation}
\item[{\bf b)}] If $\omega_1\circ \omega_2$ is a consistent path, where $\circ$ denotes the usual concatenation of paths, then
\begin{equation*}\label{prob}
    \rho_{E_N}(\omega_1\circ \omega_2)=\rho_{E_N}(\omega_1)\rho_{E_N\setminus{\omega^*_1}}(\omega_2).
\end{equation*}
\end{itemize}

As shown in Section \ref{backbone},  the backbone expansion \eqref{eq:backboneexp} for the two-point function on $\L_N$ is given by
\begin{equation*}\label{eq:exptpfunc}
    \langle \sigma_x\sigma_y\rangle_{\L_N} = \sum_{\omega\in C_{xy}(\L_N)}\rho_{E_N}(\omega),
\end{equation*}
where $x=(u_0,t_0)$, $y=(u, t)$ and  $C_{xy}(\L_N)$ is the set of all consistent paths $\omega$ with extremes $\partial\omega=\{x,y\}$ using edges
of $E_N$.

Let $\omega$ be a consistent path connecting $x$ to $y$. It is possible to break this path into $n+1$ ``planar" pieces $\omega_i$, and $n$
``vertical" steps $s_i$ (i.e. such that $|s_i|=1$) connecting two $d$-dimensional hyperplanes (note that there are $2s$ possibilities for the choice
of $s_i$). Namely,  we can write
\begin{equation}\label{ospli}
    \omega=\omega_1\circ s_1\circ \omega_2 \circ s_2\circ ...\circ s_n\circ \omega_{n+1}.
\end{equation}

We are denoting by $\o_1$ the initial piece of  the path $\o$ all contained in $\L_N^{t_0}$. This initial piece $\o_1$ of  the path $\o$ is  a
``planar" path connecting  the site
 $(u_0,t_0)$ to the  site $(u_1,t_0)$, which is the last site of $\L_N^{t_0}$ visited by  $\omega$ before leaving $\Lambda_N^{t_0}$; this path only
 uses edges of $E_N^{t_0}$. Then  $s_1$ is  the first vertical step, that is, the  edge connecting $(u_1,t_0)$ to $(u_1,t_1)$ (where $t_1=t_0+s_1$),
 which is the first site visited by the path $\omega$ after reaching a new hyperplane. Similarly, for each $k=1,...,n$, we denote by $\omega_k$ the
 consistent piece of $\omega$ that connects $(u_{k-1},t_{k-1})$ to $(u_k,t_{k-1})$, using only edges of $E(\L_N^{t_{k-1}})$. Here,
 $(u_{k-1},t_{k-1})$ is the first site of $\L_N^{t_{k-1}}$ visited after the last vertical step $s_{k-1}$ and $(u_{k},t_{k-1})$ is the last site of
 this hyperplane visited by $\omega$ before it makes another jump, that is, before it reaches another hyperplane.  Also, we denote by $s_k$ the
 vertical jump, that is, the single bond connecting $(u_k,t_{k-1})$ to $(u_k,t_k)$, the first site visited by the path  $\omega$ in a hyperplane
 different from $\L_N^{t_{k-1}}$.
Finally, the last piece $\omega_{n+1}$ of the path $\o$ connects $(u_{n},t_{n})$ to $(u_{n+1},t_{n})=(u,t)=y$, using only edges of $E_N^{t_n}$.
Note that $t_k=t_0+\sum_{j=1}^{k}s_j$, for any $k=1, \dots, n$. See Figure 1 for a sketch of this construction.

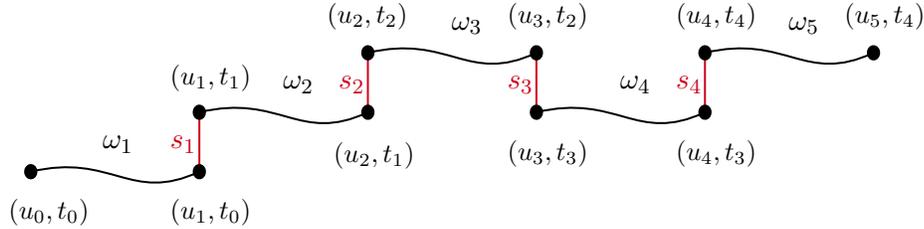
\begin{figure}[ht]\label{path}
    \centering
    \tikzset{every picture/.style={line width=0.75pt}} %set default line width to 0.75pt

\begin{tikzpicture}[x=0.75pt,y=0.75pt,yscale=-1,xscale=0.85]
%uncomment if require: \path (0,300); %set diagram left start at 0, and has height of 300

%Straight Lines [id:da3034206975450209]
\draw [color={rgb, 255:red, 208; green, 2; blue, 27 }  ,draw opacity=1 ][line width=0.75]    (140,190) -- (140,220) ;
\draw [shift={(140,220)}, rotate = 90] [color={rgb, 255:red, 208; green, 2; blue, 27 }  ,draw opacity=1 ][fill={rgb, 255:red, 208; green, 2; blue, 27 }  ,fill opacity=1 ][line width=0.75]      (0, 0) circle [x radius= 2.34, y radius= 2.34]   ;
\draw [shift={(140,190)}, rotate = 90] [color={rgb, 255:red, 208; green, 2; blue, 27 }  ,draw opacity=1 ][fill={rgb, 255:red, 208; green, 2; blue, 27 }  ,fill opacity=1 ][line width=0.75]      (0, 0) circle [x radius= 2.34, y radius= 2.34]   ;
%Curve Lines [id:da05059989979360102]
\draw    (40,220) .. controls (88.29,210.29) and (98.96,236.29) .. (140,220) ;
\draw [shift={(140,220)}, rotate = 338.35] [color={rgb, 255:red, 0; green, 0; blue, 0 }  ][fill={rgb, 255:red, 0; green, 0; blue, 0 }  ][line width=0.75]      (0, 0) circle [x radius= 3.35, y radius= 3.35]   ;
\draw [shift={(40,220)}, rotate = 348.64] [color={rgb, 255:red, 0; green, 0; blue, 0 }  ][fill={rgb, 255:red, 0; green, 0; blue, 0 }  ][line width=0.75]      (0, 0) circle [x radius= 3.35, y radius= 3.35]   ;
%Straight Lines [id:da6295940505357582]
\draw [color={rgb, 255:red, 208; green, 2; blue, 27 }  ,draw opacity=1 ][line width=0.75]    (240,160) -- (240,190) ;
\draw [shift={(240,190)}, rotate = 90] [color={rgb, 255:red, 208; green, 2; blue, 27 }  ,draw opacity=1 ][fill={rgb, 255:red, 208; green, 2; blue, 27 }  ,fill opacity=1 ][line width=0.75]      (0, 0) circle [x radius= 2.34, y radius= 2.34]   ;
\draw [shift={(240,160)}, rotate = 90] [color={rgb, 255:red, 208; green, 2; blue, 27 }  ,draw opacity=1 ][fill={rgb, 255:red, 208; green, 2; blue, 27 }  ,fill opacity=1 ][line width=0.75]      (0, 0) circle [x radius= 2.34, y radius= 2.34]   ;
%Straight Lines [id:da24467202808686173]
\draw [color={rgb, 255:red, 208; green, 2; blue, 27 }  ,draw opacity=1 ][line width=0.75]    (340,160) -- (340,190) ;
\draw [shift={(340,190)}, rotate = 90] [color={rgb, 255:red, 208; green, 2; blue, 27 }  ,draw opacity=1 ][fill={rgb, 255:red, 208; green, 2; blue, 27 }  ,fill opacity=1 ][line width=0.75]      (0, 0) circle [x radius= 2.34, y radius= 2.34]   ;
\draw [shift={(340,160)}, rotate = 90] [color={rgb, 255:red, 208; green, 2; blue, 27 }  ,draw opacity=1 ][fill={rgb, 255:red, 208; green, 2; blue, 27 }  ,fill opacity=1 ][line width=0.75]      (0, 0) circle [x radius= 2.34, y radius= 2.34]   ;
%Straight Lines [id:da8509063228783009]
\draw [color={rgb, 255:red, 208; green, 2; blue, 27 }  ,draw opacity=1 ][line width=0.75]    (440,160) -- (440,190) ;
\draw [shift={(440,190)}, rotate = 90] [color={rgb, 255:red, 208; green, 2; blue, 27 }  ,draw opacity=1 ][fill={rgb, 255:red, 208; green, 2; blue, 27 }  ,fill opacity=1 ][line width=0.75]      (0, 0) circle [x radius= 2.34, y radius= 2.34]   ;
\draw [shift={(440,160)}, rotate = 90] [color={rgb, 255:red, 208; green, 2; blue, 27 }  ,draw opacity=1 ][fill={rgb, 255:red, 208; green, 2; blue, 27 }  ,fill opacity=1 ][line width=0.75]      (0, 0) circle [x radius= 2.34, y radius= 2.34]   ;
%Curve Lines [id:da48731233079574143]
\draw    (140,190) .. controls (188.29,180.29) and (198.96,206.29) .. (240,190) ;
\draw [shift={(240,190)}, rotate = 338.35] [color={rgb, 255:red, 0; green, 0; blue, 0 }  ][fill={rgb, 255:red, 0; green, 0; blue, 0 }  ][line width=0.75]      (0, 0) circle [x radius= 3.35, y radius= 3.35]   ;
\draw [shift={(140,190)}, rotate = 348.64] [color={rgb, 255:red, 0; green, 0; blue, 0 }  ][fill={rgb, 255:red, 0; green, 0; blue, 0 }  ][line width=0.75]      (0, 0) circle [x radius= 3.35, y radius= 3.35]   ;
%Curve Lines [id:da31364805786830674]
\draw    (240,160) .. controls (288.29,150.29) and (298.96,176.29) .. (340,160) ;
\draw [shift={(340,160)}, rotate = 338.35] [color={rgb, 255:red, 0; green, 0; blue, 0 }  ][fill={rgb, 255:red, 0; green, 0; blue, 0 }  ][line width=0.75]      (0, 0) circle [x radius= 3.35, y radius= 3.35]   ;
\draw [shift={(240,160)}, rotate = 348.64] [color={rgb, 255:red, 0; green, 0; blue, 0 }  ][fill={rgb, 255:red, 0; green, 0; blue, 0 }  ][line width=0.75]      (0, 0) circle [x radius= 3.35, y radius= 3.35]   ;
%Curve Lines [id:da4146272624400478]
\draw    (340,190) .. controls (388.29,180.29) and (398.96,206.29) .. (440,190) ;
\draw [shift={(440,190)}, rotate = 338.35] [color={rgb, 255:red, 0; green, 0; blue, 0 }  ][fill={rgb, 255:red, 0; green, 0; blue, 0 }  ][line width=0.75]      (0, 0) circle [x radius= 3.35, y radius= 3.35]   ;
\draw [shift={(340,190)}, rotate = 348.64] [color={rgb, 255:red, 0; green, 0; blue, 0 }  ][fill={rgb, 255:red, 0; green, 0; blue, 0 }  ][line width=0.75]      (0, 0) circle [x radius= 3.35, y radius= 3.35]   ;
%Curve Lines [id:da3790458472580178]
\draw    (440,160) .. controls (488.29,150.29) and (498.96,176.29) .. (540,160) ;
\draw [shift={(540,160)}, rotate = 338.35] [color={rgb, 255:red, 0; green, 0; blue, 0 }  ][fill={rgb, 255:red, 0; green, 0; blue, 0 }  ][line width=0.75]      (0, 0) circle [x radius= 3.35, y radius= 3.35]   ;
\draw [shift={(440,160)}, rotate = 348.64] [color={rgb, 255:red, 0; green, 0; blue, 0 }  ][fill={rgb, 255:red, 0; green, 0; blue, 0 }  ][line width=0.75]      (0, 0) circle [x radius= 3.35, y radius= 3.35]   ;

% Text Node
\draw (81,200.4) node [anchor=north west][inner sep=0.75pt]  [color={rgb, 255:red, 0; green, 0; blue, 0 }  ,opacity=1 ]  {$\omega _{1}$};
% Text Node
\draw (121,200.4) node [anchor=north west][inner sep=0.75pt]  [color={rgb, 255:red, 208; green, 2; blue, 27 }  ,opacity=1 ]  {$s_{1}$};
% Text Node
\draw (25,232.4) node [anchor=north west][inner sep=0.75pt]  [font=\small]  {$( u_0,t_0)$};
% Text Node
\draw (121,232.4) node [anchor=north west][inner sep=0.75pt]  [font=\small]  {$( u_{1} ,t_0)$};
% Text Node
\draw (188,170.4) node [anchor=north west][inner sep=0.75pt]  [color={rgb, 255:red, 0; green, 0; blue, 0 }  ,opacity=1 ]  {$\omega _{2}$};
% Text Node
\draw (221,170.4) node [anchor=north west][inner sep=0.75pt]  [color={rgb, 255:red, 208; green, 2; blue, 27 }  ,opacity=1 ]  {$s_{2}$};
% Text Node
\draw (121,164.4) node [anchor=north west][inner sep=0.75pt]  [font=\small]  {$( u_{1} ,t_{1})$};
% Text Node
\draw (218,203.4) node [anchor=north west][inner sep=0.75pt]  [font=\small]  {$( u_{2} ,t_{1})$};
% Text Node
\draw (214,133.4) node [anchor=north west][inner sep=0.75pt]  [font=\small]  {$( u_{2} ,t_{2})$};
% Text Node
\draw (321,133.4) node [anchor=north west][inner sep=0.75pt]  [font=\small]  {$( u_{3} ,t_{2})$};
% Text Node
\draw (288,140.4) node [anchor=north west][inner sep=0.75pt]  [color={rgb, 255:red, 0; green, 0; blue, 0 }  ,opacity=1 ]  {$\omega _{3}$};
% Text Node
\draw (321,202.4) node [anchor=north west][inner sep=0.75pt]  [font=\small]  {$( u_{3} ,t_{3})$};
% Text Node
\draw (421,202.4) node [anchor=north west][inner sep=0.75pt]  [font=\small]  {$( u_{4} ,t_{3})$};
% Text Node
\draw (388,170.4) node [anchor=north west][inner sep=0.75pt]  [color={rgb, 255:red, 0; green, 0; blue, 0 }  ,opacity=1 ]  {$\omega _{4}$};
% Text Node
\draw (488,140.4) node [anchor=north west][inner sep=0.75pt]  [color={rgb, 255:red, 0; green, 0; blue, 0 }  ,opacity=1 ]  {$\omega _{5}$};
% Text Node
\draw (421,133.4) node [anchor=north west][inner sep=0.75pt]  [font=\small]  {$( u_{4} ,t_{4})$};
% Text Node
\draw (521,133.4) node [anchor=north west][inner sep=0.75pt]  [font=\small]  {$( u_{5} ,t_{4})$};
% Text Node
\draw (321,170.4) node [anchor=north west][inner sep=0.75pt]  [color={rgb, 255:red, 208; green, 2; blue, 27 }  ,opacity=1 ]  {$s_{3}$};
% Text Node
\draw (421,170.4) node [anchor=north west][inner sep=0.75pt]  [color={rgb, 255:red, 208; green, 2; blue, 27 }  ,opacity=1 ]  {$s_{4}$};

\end{tikzpicture}
    \caption{A sketch of a possible consistent path $\omega$.}
\end{figure}

We stress that since $\omega$ is consistent, each one of its pieces $\omega_i$ is also consistent. Let us set $F_1=\emptyset$ and,
for $k=2,\dots, n+1$, we set  $F_k=\o_1\circ s_1\circ\dots \circ \o_{k-1}\circ s_{k-1}$ so that
$F^*_k$ is the set of  edges of $\L_N$ canceled by the steps preceding $\omega_k$ and $s_k$.

By definition, the piece $\o_k$ of the path $\o$ is in  the $d$-dimensional hypercube
$\L_N^{t_{k-1}}$. This hypercube may have already been visited by some piece $\o_i$ of the path $\o$ with $i<k-1$ (e.g. in Figure 1, $\o_4$ is in
the same hyperplane as $\o_2$). Since the path $\omega$ is consistent, $\omega_k$ must avoid edges of the set $F_k$. Therefore, $\omega_k$ is a
consistent path which is a subset of  $E_N^{t_{k-1}}\setminus F_k$, and we denote by $C_k$ the set of all such paths with these properties.

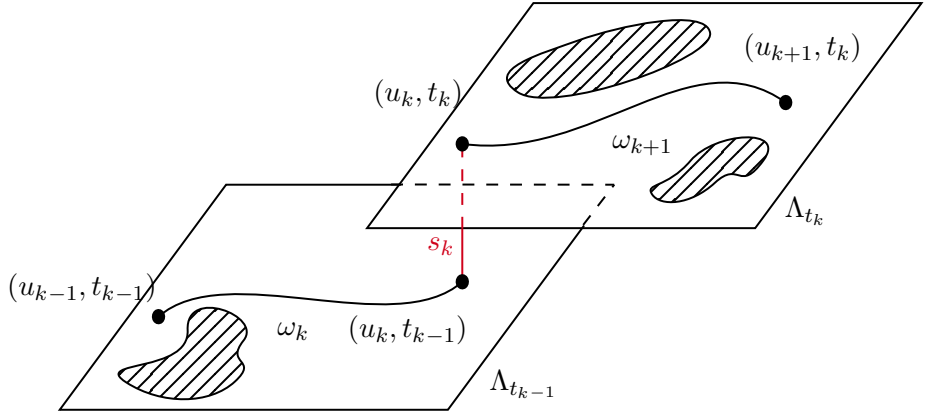
\begin{figure}[ht]
    \centering
    % Pattern Info

\tikzset{
pattern size/.store in=\mcSize,
pattern size = 5pt,
pattern thickness/.store in=\mcThickness,
pattern thickness = 0.3pt,
pattern radius/.store in=\mcRadius,
pattern radius = 1pt}
\makeatletter
\pgfutil@ifundefined{pgf@pattern@name@_ck63vne3l}{
\pgfdeclarepatternformonly[\mcThickness,\mcSize]{_ck63vne3l}
{\pgfqpoint{0pt}{0pt}}
{\pgfpoint{\mcSize+\mcThickness}{\mcSize+\mcThickness}}
{\pgfpoint{\mcSize}{\mcSize}}
{
\pgfsetcolor{\tikz@pattern@color}
\pgfsetlinewidth{\mcThickness}
\pgfpathmoveto{\pgfqpoint{0pt}{0pt}}
\pgfpathlineto{\pgfpoint{\mcSize+\mcThickness}{\mcSize+\mcThickness}}
\pgfusepath{stroke}
}}
\makeatother

% Pattern Info

\tikzset{
pattern size/.store in=\mcSize,
pattern size = 5pt,
pattern thickness/.store in=\mcThickness,
pattern thickness = 0.3pt,
pattern radius/.store in=\mcRadius,
pattern radius = 1pt}
\makeatletter
\pgfutil@ifundefined{pgf@pattern@name@_kb99c3qa1}{
\pgfdeclarepatternformonly[\mcThickness,\mcSize]{_kb99c3qa1}
{\pgfqpoint{0pt}{0pt}}
{\pgfpoint{\mcSize+\mcThickness}{\mcSize+\mcThickness}}
{\pgfpoint{\mcSize}{\mcSize}}
{
\pgfsetcolor{\tikz@pattern@color}
\pgfsetlinewidth{\mcThickness}
\pgfpathmoveto{\pgfqpoint{0pt}{0pt}}
\pgfpathlineto{\pgfpoint{\mcSize+\mcThickness}{\mcSize+\mcThickness}}
\pgfusepath{stroke}
}}
\makeatother

% Pattern Info

\tikzset{
pattern size/.store in=\mcSize,
pattern size = 5pt,
pattern thickness/.store in=\mcThickness,
pattern thickness = 0.3pt,
pattern radius/.store in=\mcRadius,
pattern radius = 1pt}
\makeatletter
\pgfutil@ifundefined{pgf@pattern@name@_mk1t2uifl}{
\pgfdeclarepatternformonly[\mcThickness,\mcSize]{_mk1t2uifl}
{\pgfqpoint{0pt}{0pt}}
{\pgfpoint{\mcSize+\mcThickness}{\mcSize+\mcThickness}}
{\pgfpoint{\mcSize}{\mcSize}}
{
\pgfsetcolor{\tikz@pattern@color}
\pgfsetlinewidth{\mcThickness}
\pgfpathmoveto{\pgfqpoint{0pt}{0pt}}
\pgfpathlineto{\pgfpoint{\mcSize+\mcThickness}{\mcSize+\mcThickness}}
\pgfusepath{stroke}
}}
\makeatother

% Pattern Info

\tikzset{
pattern size/.store in=\mcSize,
pattern size = 5pt,
pattern thickness/.store in=\mcThickness,
pattern thickness = 0.3pt,
pattern radius/.store in=\mcRadius,
pattern radius = 1pt}
\makeatletter
\pgfutil@ifundefined{pgf@pattern@name@_1e0mcuk6h}{
\pgfdeclarepatternformonly[\mcThickness,\mcSize]{_1e0mcuk6h}
{\pgfqpoint{0pt}{0pt}}
{\pgfpoint{\mcSize+\mcThickness}{\mcSize+\mcThickness}}
{\pgfpoint{\mcSize}{\mcSize}}
{
\pgfsetcolor{\tikz@pattern@color}
\pgfsetlinewidth{\mcThickness}
\pgfpathmoveto{\pgfqpoint{0pt}{0pt}}
\pgfpathlineto{\pgfpoint{\mcSize+\mcThickness}{\mcSize+\mcThickness}}
\pgfusepath{stroke}
}}
\makeatother
\tikzset{every picture/.style={line width=0.75pt}} %set default line width to 0.75pt

\begin{tikzpicture}[x=0.75pt,y=0.75pt,yscale=-1,xscale=0.85]
%uncomment if require: \path (0,300); %set diagram left start at 0, and has height of 300

%Shape: Parallelogram [id:dp44192087953421755]
\draw   (392.03,59.33) -- (622.33,59.33) -- (523.63,172.93) -- (293.33,172.93) -- cycle ;
%Shape: Parallelogram [id:dp160647840911881]
\draw   (209.7,151) -- (440,151) -- (341.3,264.6) -- (111,264.6) -- cycle ;
%Straight Lines [id:da7323161107397018]
\draw [color={rgb, 255:red, 255; green, 255; blue, 255 }  ,draw opacity=1 ][line width=2.25]    (313.6,151.68) -- (448.53,150.8) ;
%Straight Lines [id:da20446352268140355]
\draw [color={rgb, 255:red, 255; green, 255; blue, 255 }  ,draw opacity=1 ][line width=2.25]    (440,151) -- (421.6,171.28) ;
%Straight Lines [id:da9745602021344626]
\draw  [dash pattern={on 4.5pt off 4.5pt}]  (307.6,150.88) -- (440,151) ;
%Straight Lines [id:da11101106365532765]
\draw  [dash pattern={on 4.5pt off 4.5pt}]  (421.6,171.28) -- (440,151) ;
%Straight Lines [id:da5061282615458114]
\draw [color={rgb, 255:red, 208; green, 2; blue, 27 }  ,draw opacity=1 ] [dash pattern={on 4.5pt off 4.5pt}]  (349.84,173.2) -- (349.84,130.4) ;
%Straight Lines [id:da35404940042560673]
\draw [color={rgb, 255:red, 208; green, 2; blue, 27 }  ,draw opacity=1 ]   (349.84,173.2) -- (349.84,200) ;
%Shape: Boxed Bezier Curve [id:dp9029100893743001]
\draw    (349.84,130.4) .. controls (427.84,139.4) and (485.73,75.53) .. (541.73,109.53) ;
\draw [shift={(541.73,109.53)}, rotate = 31.26] [color={rgb, 255:red, 0; green, 0; blue, 0 }  ][fill={rgb, 255:red, 0; green, 0; blue, 0 }  ][line width=0.75]      (0, 0) circle [x radius= 3.35, y radius= 3.35]   ;
\draw [shift={(349.84,130.4)}, rotate = 6.58] [color={rgb, 255:red, 0; green, 0; blue, 0 }  ][fill={rgb, 255:red, 0; green, 0; blue, 0 }  ][line width=0.75]      (0, 0) circle [x radius= 3.35, y radius= 3.35]   ;
%Curve Lines [id:da12385104463058827]
\draw    (169.6,217.6) .. controls (209.6,187.6) and (309.84,230) .. (349.84,200) ;
\draw [shift={(349.84,200)}, rotate = 323.13] [color={rgb, 255:red, 0; green, 0; blue, 0 }  ][fill={rgb, 255:red, 0; green, 0; blue, 0 }  ][line width=0.75]      (0, 0) circle [x radius= 3.35, y radius= 3.35]   ;
\draw [shift={(169.6,217.6)}, rotate = 323.13] [color={rgb, 255:red, 0; green, 0; blue, 0 }  ][fill={rgb, 255:red, 0; green, 0; blue, 0 }  ][line width=0.75]      (0, 0) circle [x radius= 3.35, y radius= 3.35]   ;
%Shape: Polygon Curved [id:ds8057685378302972]
%\draw  [pattern=_ck63vne3l,pattern size=6pt,pattern thickness=0.75pt,pattern radius=0pt, pattern color={rgb, 255:red, 0; green, 0; blue, 0}] (244.88,162) .. controls (267.28,155.16) and (289.68,154.8) .. (278.08,167.2) .. controls (266.48,179.6) and (267.68,167.6) .. (252.88,179.2) .. controls (238.08,190.8) and (229.68,199.96) .. (208.48,188.4) .. controls (187.28,176.84) and (222.48,168.84) .. (244.88,162) -- cycle ;
%%Shape: Polygon Curved [id:ds026357410405794868]
\draw  [pattern=_kb99c3qa1,pattern size=6pt,pattern thickness=0.75pt,pattern radius=0pt, pattern color={rgb, 255:red, 0; green, 0; blue, 0}] (186.88,224.4) .. controls (185.68,202.4) and (231.68,217.2) .. (220.08,229.6) .. controls (208.48,242) and (229.68,238.8) .. (214.88,250.4) .. controls (200.08,262) and (171.68,262.36) .. (150.48,250.8) .. controls (129.28,239.24) and (188.08,246.4) .. (186.88,224.4) -- cycle ;
%Shape: Polygon Curved [id:ds9976969289803401]
\draw  [pattern=_mk1t2uifl,pattern size=6pt,pattern thickness=0.75pt,pattern radius=0pt, pattern color={rgb, 255:red, 0; green, 0; blue, 0}] (419.28,78) .. controls (441.68,71.16) and (494.48,61.2) .. (496.88,74.4) .. controls (499.28,87.6) and (404.08,115.96) .. (382.88,104.4) .. controls (361.68,92.84) and (396.88,84.84) .. (419.28,78) -- cycle ;
%Shape: Polygon Curved [id:ds27379753631558357]
\draw  [pattern=_1e0mcuk6h,pattern size=6pt,pattern thickness=0.75pt,pattern radius=0pt, pattern color={rgb, 255:red, 0; green, 0; blue, 0}] (485.36,138.8) .. controls (499.76,124.4) and (542.96,122) .. (528.48,138.8) .. controls (514,155.6) and (514.56,138.4) .. (499.76,150) .. controls (484.96,161.6) and (467.36,162) .. (462.56,156) .. controls (457.76,150) and (470.96,153.2) .. (485.36,138.8) -- cycle ;

% Text Node
\draw (238.5,219) node [anchor=north west][inner sep=0.75pt]    {$\omega _{k}$};
% Text Node
\draw (327.33,176.27) node [anchor=north west][inner sep=0.75pt]  [color={rgb, 255:red, 208; green, 2; blue, 27 }  ,opacity=1 ]  {$s_{k}$};
% Text Node
\draw (437.13,124.73) node [anchor=north west][inner sep=0.75pt]    {$\omega _{k+1}$};
% Text Node
\draw (78,194.4) node [anchor=north west][inner sep=0.75pt]    {$( u_{k}{}_{-1} ,t_{k}{}_{-1})$};
% Text Node
\draw (280.2,215.6) node [anchor=north west][inner sep=0.75pt]    {$( u_{k}{} ,t_{k}{}_{-1})$};
% Text Node
\draw (295.33,96.53) node [anchor=north west][inner sep=0.75pt]    {$( u_{k} ,t_{k})$};
% Text Node
\draw (514.93,73.13) node [anchor=north west][inner sep=0.75pt]    {$( u_{k+1}{} ,t_{k})$};
% Text Node
\draw (364.16,243.24) node [anchor=north west][inner sep=0.75pt]    {$\Lambda _{t_{k}{}_{-1}}$};
% Text Node
\draw (539.36,156.04) node [anchor=north west][inner sep=0.75pt]    {$\Lambda _{t_{k}}$};
\end{tikzpicture} 
     \caption{A sketch of a transition between hyperplanes.}
\end{figure}

Now, for $n=0,1,2,\dots $, set $\mathcal{U}_n=(u_1,...,u_{n+1})$  and  $\mathcal{S}_n=(s_1,...,s_n)$, with the convention that
$\mathcal{S}_0=\emptyset$.
Then, we can write
$$
\langle \sigma_x\sigma_y\rangle_{\L_N}=\sum_{n\ge 0}\sum_{\substack{\mathcal{U}_n,\mathcal{S}_n \\ u_{n+1}=u \\ t_n=t}}\sum_{\omega_1\in
C_1}...\sum_{\omega_{n+1}\in C_{n+1}}\rho_{E_N}(\o),
$$
with $\o$ given by \eqref{ospli}.
Summing over $y\in\Lambda_N$,  we get
\begin{equation}\label{eq:integralpt1}
    \sum_{y\in \L_N}\langle \sigma_x\sigma_y\rangle_{\L_N}  = \sum_{n\ge 0}\sum_{\mathcal{U}_n,\mathcal{S}_n}\sum_{\omega_1\in
    C_1}...\sum_{\omega_{n+1}\in C_{n+1}}\rho_{E_N}(\omega).
\end{equation}
%Where in the last equality, we are using the sum over $x=(u,t)=(u_{n+1},t_n)$ to allow $u_{n+1}$ and $t_n$ to run free.
By Property {\bf b)} given at the beginning of this section, we get
\begin{equation*}\label{eq:rhodec}
    \rho_{E_N}(\omega)=\rho_{E_N-F_{n+1}}(\omega_{n+1})\prod_{k=1}^{n}\rho_{E_N-F_k}(\omega_k)\rho_{E_N-F^*_k}(s_k),
\end{equation*}
where we recall $F_1\coloneqq \varnothing$.

Now, using the bound \eqref{prot}, and recalling that $s_k$ is a single vertical edge (and thus with $J_{s_k}=J_s$),  it holds that
$$
\rho_{E_N-F^*_k}(s_k)\le \tanh( J_s),
$$
for any $k=1,\cdots,n$. Therefore,
\begin{equation*}\label{eq:rhodecbound}
    \rho_{E_N}(\omega)\leq [\tanh( J_s)]^n\prod_{k=1}^{n+1}\left[\rho_{E_N- F_k}(\omega_k)\right].
\end{equation*}
Observe that $E_N^{t_{k-1}}-F_k\subset E_N- F_k$. Moreover,   since $\omega$ is consistent, $\omega_k$ only  uses  edges of $E_N^{t_{k-1}}-F_k$. Hence, we can apply Property {\bf a)} and inequality \eqref{proa} to obtain
$$
\rho_{E_N- F_k}(\omega_k)\le \rho_{E_N^{t_{k-1}}- F_k}(\omega_k),
$$
for any $k=1\cdots,n+1$, yielding
\begin{equation}\label{eq:rhodecbound2}
    \rho_{E_N}(\omega)\leq[\tanh(J_s)]^n\prod_{k=1}^{n+1}
    \rho_{E_N^{t_{k-1}}- F_k}(\omega_k).
\end{equation}
Plugging \eqref{eq:rhodecbound2} in \eqref{eq:integralpt1}, we write

\begin{equation*}\label{eq:boundchi1}
    \sum_{y\in \L_N}\langle \sigma_x\sigma_y\rangle_{\L_N} \leq \sum_{n\ge 0}
    [\tanh( J_s)]^n\sum_{\mathcal{U}_n,\mathcal{S}_n}S_1\cdot ...\cdot S_{n+1},
\end{equation*}
where, for $k=1,...,n+1$, we have set
$$
S_k=S_k(u_{k-1},u_k, t_{k-1}) = \sum_{\omega_k\in C_{k}}\rho_{\L_N^{t_{k-1}}- F_k}(\omega_k).
$$
Then, we can write
$$
\sum_{\mathcal{U}_n,\mathcal{S}_n}S_1\cdot ...\cdot S_{n+1}=\sum_{u_1} S_1\sum_{s_1}\sum_{u_2}S_2\dots
\sum_{u_{n-1}}S_{n-1}\sum_{s_{n-1}}\sum_{u_n}S_n\sum_{s_n}\sum_{u_{n+1}}S_{n+1}.
$$
Observe now that,  for any $k=1,\cdots,n$,

\begin{equation*}
\begin{split}
    S_k & = \sum_{\omega_k\in C_{k}}\rho_{\L_N^{t_{k-1}}-F_k}(\omega_k)\\ &
    =\langle\sigma_{(u_{k-1},t_{k-1})}\sigma_{(u_{k},t_{k-1})}\rangle_{\L_N^{t_{k-1}}-F_{k}}\\
    &\le \langle\sigma_{(u_{k-1},t_{k-1})}\sigma_{(u_{k},t_{k-1})}\rangle_{\L_N^{t_{k-1}}} .
\end{split}
\end{equation*}
where the last line follows by the GKS inequalities.
Therefore, for any fixed $u_1,\dots ,u_n$, and any fixed $s_1,\dots ,s_{n-1}$, we get
\begin{equation*}
    \begin{split}
        \sum_{s_n}\sum_{u_{n+1}}S_{n+1} & \le \sum_{s_n} \sum_{u_{n+1}}\langle\sigma_{(u_n,t_n)}
        \sigma_{(u_{n+1},t_n)}\rangle_{\Lambda_N^{t_{n}}}\\
        & \leq \sum_{s_n}\sup_{u\in \L_N^{t_n}}\sum_{u_{n+1}}\langle\sigma_{(u,t_n)}\sigma_{(u_{n+1},t_n)}\rangle_{\Lambda_N^{t_{n}}} \\
        & = \sum_{s_n} \chi_{\L_N^{t_n}}(J_d)\\
        & = 2s \chi_{\Lambda_N^{t_n}}(J_d)\\
        & = 2s \chi_{\Lambda_N^{0}}(J_d).\\
    \end{split}
\end{equation*}
Proceeding iteratively for the sums $ \sum_{s_{k-1}}\sum_{u_{k}}S_{k}$ (with $k=n, n-1, \dots,  1$), and with the convention that
$\sum_{s_0}=1$, we obtain
$$
\sum_{\mathcal{U}_n,\mathcal{S}_n}S_1\cdot ...\cdot S_{n+1}\le (2s)^n[\chi_{\L_N^{0}}(J_d)]^{n+1},
$$
whence, for any $x\in \L_N$,

%Similarly, for each $k=1,...,n$, summing over $s_k$ and $u_k$ in an inductive way, we get
%\begin{equation}
%    \begin{split}
%        \sum_{s_k}\sum_{u_k}S_{k} &  \leq \sum_{s_k}\sum_{u_k}\tanh(\beta
%J)\langle\sigma_{(u_{k-1},t_{k-1})}\sigma_{(u_{k},t_{k-1})}\rangle_{\Lambda_{t_{k-1}}-F_{k}}\\ &  = 2s\tanh(\beta
%J)\sum_{u_k}\langle\sigma_{(u_{k-1},t_{k-1})}\sigma_{(u_{k},t_{k-1})}\rangle_{\Lambda_{t_{k-1}-F_{k}}}\\ &  \leq 2s\tanh(\beta
%J)\sum_{u_k}\langle\sigma_{(u_{k-1},t_{k-1})}\sigma_{(u_{k},t_{k-1})}\rangle_{\Lambda_{t_{k-1}}}.
%    \end{split}
%\end{equation}
%Again, we use Griffiths's inequality in the last line. Therefore, this sum can be bounded from above by
%\begin{equation}
%    \sum_{s_k}\sum_{u_k}S_{k}\leq 2s\tanh(\beta J)\chi^{\Lambda_{t_0}}.
%\end{equation}
%Finally, returning to \eqref{eq:boundchi1}, we conclude that
\begin{equation*}
    \sum_{y\in \L_N}\langle \sigma_x\sigma_y\rangle_{\L_N}\leq \sum_{n\ge 0}(2s\tanh( J_s))^{n}[\chi_{\L_N^{0}}(J_d)]^{n+1}.
\end{equation*}
Finally, taking the limit $N\to\infty$ and recalling the definitions of $\chi_{d+s}(J_d,J_s)$ and $\chi_{d}(J_d)$ given in \eqref{kids} and
\eqref{kid}, respectively, we get
$$
\chi_{d+s}(J_d,J_s)\le   \sum_{n\ge 0}(2s\tanh( J_s))^{n}[\chi_{d}(J_d)]^{n+1}.
$$
The r.h.s. of the inequality above is finite provided that
$$
\tanh(J_s)< \frac{1}{2s\chi_{d}(J_d)},
$$
and thus the proof of Theorem \ref{th1} is concluded.

\section*{Acknowledgements}
Estevão Borel was partially supported by Fundação Coordenação de Aperfeiçoamento de Pessoal de Nível Superior (CAPES). Aldo Procacci has been
partially supported by the Brazilian science foundations Conselho Nacional de Desenvolvimento Científico e Tecnológico (CNPq), CAPES and Fundação de
Amparo à Pesquisa do Estado de Minas Gerais (FAPEMIG). Rémy Sanchis was supported by CNPq, and by FAPEMIG, grants APQ-00868-21 and RED-00133-21.
Roger Silva was partially supported by FAPEMIG, grant APQ-00774-21.

\section*{Conflict of interest statement}

The authors have no conflicts of interest to declare.
All co-authors have seen and agree with the contents of the manuscript and
there is no financial interest to report. We certify that the submission is
original work and is not under review at any other publication.

\section*{Data availability statement}

Data sharing not applicable to this article as no datasets were generated or analysed during the current study.


\begin{thebibliography}{99}

\bibitem{Abe} Abe R.: Some remarks on pertubation theory and phase transition with an application to anisotropic Ising model,  \emph{Prog. Theor.
    Phys.}, {\bf 44}, 339--347, (1970).



\bibitem{A} Aizenman M.: Geometric analysis of $\psi^4$
 fields and Ising models. I, II,  \emph{Commun. Math. Phys.}, {\bf 86}, 1--48, (1982).

 \bibitem{AD} Aizenman, M., Duminil-Copin, H.: Marginal triviality of the scaling limits of
critical 4D Ising and $\phi^4_4$ models. \emph{Ann. Math}, {\bf 194}, no. 1, 163–235  (2021).

\bibitem{ADS}  Aizenman M., Duminil-Copin H., and Sidoravicius V.: Random Currents and Continuity of Ising Model’s
 Spontaneous Magnetization, \emph{Commun. Math. Phys.}, {\bf 334}, 719–742, (2015).

 \bibitem{ADTW} Aizenman, M., Duminil-Copin, H.,  Tassion, V., Warzel, S.:  Emergent planarity in two-dimensional Ising models with finite-range
     interactions, \emph{Invent. Math.}, {\bf 216}, no. 3, 661–743 (2019).]

\bibitem{AF} Aizenman M. and Fernández R.: On the critical behavior of the magnetization in high-dimensional Ising models,  \emph{J. Stat. Phys.},
    {\bf 44}, 393--454, (1986).

\bibitem{AG} L A Aviles and R P Gammag:  \emph{J. Phys.}: Conf. Ser. 2543 012006, (2023).

%\bibitem{CGN} Camia F., Garban C. and Newman C.M: The Ising magnetization exponent on $\mathbb{Z}^2$ is 1/15,  %\emph{Probab. Theory Relat.
%Fields}, {\bf 160}, (2014), 175--187.

\bibitem{D} Duminil-Copin, H.: Random currents expansion of the Ising model, arXiv:1607.06933 (2016).

\bibitem{FSBB} D. Farsal; M. Snina1; M. Badia;  M. Bennai: Critical Properties of Two-dimensional Anisotropic Ising
Model on a Square Lattice, \emph{J Supercond. Nov. Magn.}, {\bf 30}, 2187–2195, (2017).

\bibitem{F} Fisher M.E.: Critical Temperatures of Anisotropic Ising Lattices. II. General Upper Bounds, \emph{Phys. Rev.}, {\bf 162}, 480 (1967).

%\bibitem{FV} Friedli S. and Velenik Y.: \emph{Statistical mechanics of lattice systems}, Cambridge University $Press, (2018).

\bibitem{KKB}  Kamiya Y., Kawashima1 N. and Batista C.D.: Dimensional crossover in the quasi-two-dimensional Ising-O(3) model,  \emph{J. Phys.}:
    Conf. Ser. 320 012023, (2011).
\bibitem{KASL} Kim Y.C., Anisimov M.A., Sengers J.V. and Luijten E.: Crossover Critical Behavior in the Three-Dimensional Ising Model, \emph{J. of
    Stat. Phys.}, {\bf 110}, 591–609, (2003).

\bibitem{L} Lee K. W.: Dimensional Crossover in the anisotropic 3D Ising model: a Monte Carlo study,   \emph{J. of the Korean Phys. Soc.}, {\bf 40},
    L398--L401, (2002).

\bibitem{LS1} Liu L. L. and Stanley H. E.: Some results concerning the crossover behavior of Quasi-two-dimensional and quasi-one-dimensional
    systems,   \emph{Phy. Review Lett.}, {\bf 29}, 927--930, (1972).

\bibitem{LS2} Liu L. L. and Stanley H. E.: Quasi-one-dimensional and quasi-two-dimensional magnetic systems: determination of crossover temperature
    and scaling with anisotropy parameters,   \emph{Phy. Review B}, {\bf 8}, 2279--2299, (1973).


\bibitem{MPS} Mazel A., Procacci A. and  Scoppola B.: Gas Phase of Asymmetric Nearest Neighbor Ising Model. \emph{J. of Stat. Phys.}, {\bf 106},
    1241–1248, (2002).

\bibitem{NJ} Navarro R. and de Jongh L.J.: On the lattice-dimensionality crossovers in magnetic Ising systems, \emph{Physica 94B}, 67-77, (1978).

%Model: Again: Ising $3d$ with constant $J_{xy}$ in the xy plane and $J_z$ in the z direction with the ratio $J_{xy}/J_z$.

%\bibitem{O} Onsager L.: Crystal Statistics. I. A Two-dimensional model with an order-disorder transition,   %\emph{Phy. Review}, {\bf 65}, (1944),
117--149.

\bibitem{OE}  Oitmaa J. and Enting I.G.: Critical behaviour of the anisotropic Ising model, \emph{Physics Letters},
{\bf 36} A, number 2, (1971)

\bibitem{SS}  Sanchis R. and Silva, R. W. C.: Dimensional Crossover in Anisotropic Percolation on
 $\mathbb Z^{d+s}$, \emph{J. Stat. Phys.}, {\bf 169}, 981–988, (2017).

%\bibitem{S} Sokal A.D.: A rigorous inequality for the specific heat of an Ising or $\phi^4$ ferromagnet,   %\emph{Phy. Letters}, {\bf 71}, (1979),
451--453.


\bibitem{Su} Suzuki M.: Scaling with a parameter in spin systems near the critical point. I,   \emph{Prog. Theor. Phys.}, {\bf 46}, 1054--1070,
    (1971).

\bibitem{Ya} Yamagata A.: Finite-size effects in the quasi-two-dimensional Ising model, \emph{Physica A},
{\bf 205}, Issue 4, 665-676, (1994)


\bibitem{Y} Yang C.N.: The spontaneous magnetization on a two-dimensional Ising model,  \emph{Phy. Review}, {\bf 85}, 808--816, (1952).

\bibitem{Yu}  Yurishchev M. A.: Lower and Upper Bounds on the Critical Temperature
 for Anisotropic Three-Dimensional Ising Model, \emph{Journal of Experimental and Theoretical Physics}, {\bf 98}, No. 6, pp. 1183–1197, (2004).

\bibitem{VPRL}  Viswanathan G.M., Portillo M.A.G.,  Raposo E.P. and da Luz M.H.E.: What Does It Take to Solve
 the 3D Ising Model? Minimal Necessary Conditions for a Valid Solution,  \emph{Entropy},  {\bf 24}, 1665, (2022).


\bibitem{ZSH} Zandvliet H.J.W., Saedi A. and Hoede C.: The anisotropic 3D Ising model, \emph{Phase Transitions}, {\bf 80}, 981-986, (2007).

\end{thebibliography}
\end{document}